# A New Method of Measuring $^{81}$Kr and $^{85}$Kr Abundances in Environmental Samples

X. Du,[1,2] R. Purtschert,[3] K. Bailey,[1] B. E. Lehmann,[3] R. Lorenzo,[3] Z.-T. Lu,[1] P. Mueller,[1] T. P. O'Connor,[1] N. C. Sturchio,[4] L. Young[5]

[1]*Physics Division, Argonne National Laboratory, Argonne, USA;*

[2]*Physics Department, Northwestern University, Evanston, USA;*

[3]*Climate and Environmental Physics, Physics Institute, University of Bern, Switzerland;*

[4]*Department of Earth and Environmental Sciences, University of Illinois at Chicago, Chicago, USA;*

[5]*Chemistry Division, Argonne National Laboratory, Argonne, USA*

**Abstract.** We demonstrate a new method for determining the $^{81}$Kr/Kr ratio in environmental samples based upon two measurements: the $^{85}$Kr/$^{81}$Kr ratio measured by Atom Trap Trace Analysis (ATTA) and the $^{85}$Kr/Kr ratio measured by Low-Level Counting (LLC). This method can be used to determine the mean residence time of groundwater in the range of $10^5 - 10^6$ a. It requires a sample of 100 μl STP of Kr extracted from approximately two tons of water. With modern atmospheric Kr samples, we demonstrate that the ratios measured by ATTA and LLC are directly proportional to each other within the measurement error of ±10%; we calibrate the $^{81}$Kr/Kr ratio of modern air measured using this method; and we show that the $^{81}$Kr/Kr ratios of samples extracted from air before and after the development of the nuclear industry are identical within the measurement error.

$^{81}$Kr ($t_{1/2}$ = 2.3 ×10$^5$ a, $^{81}$Kr/Kr ~ 10$^{-13}$) has been proposed as the ideal tracer isotope for dating old water and ice in the age range of 10$^5$-10$^6$ a (Loosli and Oeschger, 1969). $^{81}$Kr is mainly produced in the upper atmosphere by cosmic-ray induced spallation and neutron activation of stable krypton. Because of the constancy of the cosmic ray flux and the fact that the atmosphere is well-mixed and represents the only significant terrestrial Kr reservoir, the $^{81}$Kr abundance in the atmosphere is expected to be constant on the time scale of its lifetime. Subsurface sources and sinks for $^{81}$Kr other than radioactive decay are most likely negligible (Lehmann et al., 2003). Human activities involving nuclear fission have a negligible effect on the $^{81}$Kr concentration because its direct yield from spontaneous fission of $^{238}$U is small and because the stable $^{81}$Br shields $^{81}$Kr from the decay of other fission products. On the other hand, $^{85}$Kr ($t_{1/2}$ = 10.8 a, $^{85}$Kr/Kr ~ 10$^{-11}$) is a fission product of $^{235}$U and $^{239}$Pu, and is released into the atmosphere primarily by nuclear fuel reprocessing. Its abundance has increased by six orders of magnitude since the 1950's. $^{85}$Kr can be used as a tracer to study air and ocean currents, determine residence time of young groundwater in shallow aquifers, and monitor nuclear-fuel processing activities (Loosli, 1992).

For $^{85}$Kr analysis, Low-Level Counting (LLC) is performed routinely in several specialized laboratories around the world (Loosli, 1992). LLC was the first method used to detect $^{81}$Kr and measure its abundance in the atmosphere (Loosli, 1969).

However, LLC is too inefficient for practical analysis of $^{81}$Kr because only a fraction $3 \times 10^{-8}$ of $^{81}$Kr atoms in a sample decay during a typical 100 hour measurement. In general, counting atoms is preferable to counting decays for analysis of long-lived isotopes such as $^{81}$Kr. Two laser-based methods, Resonance Ionization Mass Spectrometry (RIMS) and Photon Burst Mass Spectrometry (PBMS), have both been developed towards detecting $^{81}$Kr and $^{85}$Kr in environmental samples (Lu and Wendt, 2003). Using RIMS, Lehmann et al. (1991) reported the measurement of $^{81}$Kr in krypton gas extracted from water samples in the Milk River aquifer in Canada. Moreover, in recent work based on Accelerator Mass Spectrometry (AMS), using a high-energy ($\sim$ 4 GeV) accelerator in order to separate $^{81}$Kr from the abundant $^{81}$Br, Collon et al. (2000) analyzed $^{81}$Kr in old groundwater samples from Australia at an efficiency of $2 \times 10^{-5}$, with which a 10% measurement required 500 μl STP krypton extracted from 16 tons of groundwater.

Atom Trap Trace Analysis (ATTA) is a relatively new atom-counting method, and was used to detect both $^{81}$Kr and $^{85}$Kr in natural atmospheric samples (Chen et al., 1999). It uses a table-top apparatus in a standard laboratory environment. In ATTA, an atom of a particular isotope is selectively captured by a magneto-optical trap (MOT) and detected by observing its fluorescence. When the laser frequency is tuned to within a few natural linewidths on the low-frequency side of the resonance of the desired isotope, only atoms of this particular isotope are trapped. Atoms of other isotopes are either deflected before reaching the trap or are allowed to pass through the trap without being captured. An atom can be trapped and observed for 100 ms or longer, during which $10^6$ fluorescence photons can be induced from a single trapped atom and as many as $10^4$ photons can be detected, thereby allowing the counting of single atoms to be done with a high signal-to-noise ratio as well as a superb selectivity. Indeed ATTA is immune to interference from other isotopes, elements, or molecules. Since 1999, system improvements including a more intense source of metastable Kr atoms, a recirculating vacuum system and better optical arrangements have been implemented. At present, the count rate of $^{81}$Kr in a sample of modern atmospheric Kr is 12 hr$^{-1}$, the counting efficiency is $1 \times 10^{-4}$, and the required sample size is 50 μl STP Kr.

For accurate analysis, the $^{81}$Kr count rate is normalized to that of a control isotope, and the $^{81}$Kr abundance is determined by the ratio of the two count rates. By frequently switching the system between counting $^{81}$Kr and counting the control isotope, much of the common-mode noise on count rates due to the variation of the instrumental parameters is cancelled in the ratio. Furthermore, age is determined by the ratio of the $^{81}$Kr abundance in the sample over that in the atmosphere; it is in effect a ratio of ratios, which further cancels any systematic effects. In principle, the stable $^{83}$Kr can be used as the control isotope. In practice, however, its count rate by ATTA, $10^9$ s$^{-1}$, is too high for atom counting. On the other hand, $^{85}$Kr at the $10^{-11}$ abundance level can be introduced into the sample and calibrated accurately with LLC. The spiked $^{85}$Kr can then be counted at the single atom level and serve as a reliable control isotope. On the other hand, in young (<100 a) atmospheric Kr samples, the natural $^{81}$Kr can be used as the control isotope to measure variations in the $^{85}$Kr abundance of unspiked samples.

In order to demonstrate the validity of ATTA for quantitative analysis, we measured the $^{85}$Kr/$^{81}$Kr ratios of ten Kr samples using ATTA at Argonne. Among these Kr samples extracted from modern air (age <100 a), the $^{81}$Kr/Kr ratios are expected to be identical. On the other hand, the $^{85}$Kr/Kr ratios are expected to vary and were measured using LLC at Bern. The $^{85}$Kr/$^{81}$Kr values measured with ATTA are then compared with the $^{85}$Kr/Kr values obtained with LLC for the following purposes: 1) to verify the proportional relationship between the two sets of ratios; 2) to calibrate the $^{81}$Kr/Kr ratio of modern air measured by ATTA, which serves as the initial ratio in the calculation of groundwater residence times; 3) to compare the $^{81}$Kr/Kr ratios of atmospheric Kr samples extracted before and after the development of nuclear industry, which are often referred to as pre-bomb and post-bomb samples.

**Low-Level Counting (LLC)**

At the LLC laboratory of the University of Bern, $^{85}$Kr was measured with gas proportional counting (Fig. 1). 99.6 % of the $^{85}$Kr atoms β-decay with a maximum energy of 0.69 MeV. Because of this relatively high β-energy and in order to identify gas impurities the whole energy spectrum above 4 keV is measured using a multi-channel analyzer. The gas multiplication factor is calibrated every other day using an external $^{241}$Am source which induces a fixed 8.5 keV x-ray line. The energy deposition within the gas volume increases with gas pressure in the counter. Therefore counters of 16 cm$^3$ and 22 cm$^3$ are operated at 2-5 bars using P-10 gas (90% argon, 10% methane, and krypton sample). The Kr abundance in the counting gas is determined from the filling pressure and the Kr/Ar ratio measured for each sample by mass spectrometry after counting. In order to achieve very low background count-rates, counting is carried out in a laboratory 35 meters below the earth surface (shielding ~ 70 m water equivalent). The walls of the laboratory were built with special low-activity concrete. In addition, old lead shielding surrounds the counters that are made from high-purity copper and are operated in anticoincidence arrangements inside guard counters (Forster et al., 1992). The remaining background count-rate for a 16 cc counter filled with a pressure of 2.3 bar is 1.2 hr$^{-1}$ compared to a signal of approximately 40 hr$^{-1}$ for a modern Kr sample. This corresponds to a detector efficiency of 70%. Typical counting times are 3-6 days, during which (0.5 – 1) × 10$^{-3}$ of the $^{85}$Kr atoms in the sample decay. The detection limit is about 0.01 Bq/cc Kr with a minimum Kr sample size of 5 μl.

The activity of $^{85}$Kr in the atmosphere has been steadily increasing, from a pre-bomb level of ~ 10$^{-6}$ Bq/cc Kr, to a present-day level of 1.4 Bq/cc Kr (Loosli et al., 2000), corresponding to a $^{85}$Kr/Kr ratio of 2.5 × 10$^{-11}$. Pre-bomb samples extracted from old groundwater have negligible amount of $^{85}$Kr. For ATTA measurements, in which $^{85}$Kr is used as a control isotope, a calibrated amount of $^{85}$Kr are added to the samples. The spike gas should have a high $^{85}$Kr/$^{81}$Kr ratio in order to minimize the portion of $^{81}$Kr added to the samples, especially for very old samples with low $^{81}$Kr/Kr. In this work, the gas used for spiking the pre-bomb samples has a

$^{85}$Kr/Kr ratio of 1.02 × 10$^{-9}$ and was produced by diluting a 1.85 × 10$^{8}$ Bq $^{85}$Kr standard. Ten samples were prepared and analyzed (Table 1). The $^{85}$Kr/Kr ratios of samples 1, 2, 3, 5, 6 and 9 were measured by LLC; the $^{85}$Kr/Kr ratios of the mixtures (samples 4, 7, 8 and 10) were calculated according to the mixing ratios.

**Atom Trap Trace Analysis (ATTA)**

The principle of ATTA and the design of a first-generation ATTA system have been described elsewhere (Chen et al., 1999). Here we briefly describe the design and operation of a second-generation ATTA system (Fig. 2). An all-diode-laser system supplies the laser beams for transverse cooling, slowing, and trapping. The frequency of the laser, after an approximate 800 MHz offset generated by a tunable acousto-optic modulator (AOM) to compensate for isotope shifts, is locked to the resonance of the $5s[3/2]_2$ → $5p[5/2]_3$ transition of the abundant $^{84}$Kr in a reference vapor cell. The electronic control of the AOM offset frequency allows us to tune the laser frequencies to match the transitions of $^{83}$Kr, $^{85}$Kr, and $^{81}$Kr, and to trap atoms of these isotopes, respectively. The Kr sample is injected through a leak valve into the source chamber where a rf-driven discharge produces Kr atoms in the metastable $5s[3/2]_2$ level. The emerging beam of metastable Kr atoms is cooled in both transverse directions to enhance its flux in the forward direction by a factor of 20. The metastable Kr atoms then enter a 1.2 meter long Zeeman slower, where they are decelerated to 20 m/s as they enter the trap chamber. In the trap chamber, a MOT is used to capture the slow atoms and confine them in a sub-millimeter region in the center of the chamber. A single trapped Kr atom scatters photons at a rate of $10^7$ s$^{-1}$, of which 5% are collected and imaged onto an avalanche photo-diode with a photon-counting efficiency of 25% at 811 nm wavelength. The resulting single-atom signal is a photon count rate of 20 kHz above a background of 13 kHz (Fig. 3). In a 60 ms counting period, the single atom detection achieves a signal-to-noise ratio of 45. The vacuum system is differentially pumped by three turbo pumps so as to maintain a pressure of a few mTorr in the source chamber and a pressure of ~ 10$^{-8}$ Torr in the trap chamber. A getter pump in the source chamber removes reactive gases, such as hydrogen, water, etc., from the vacuum system while leaving noble gases including the Kr sample intact. A novel feature of the system is that it can be switched into a mode which repeatedly circulates the Kr atoms through the vacuum system and thereby improves the counting efficiency by a factor of a thousand to reach 1 × 10$^{-4}$. For an analysis, the system is filled to the operation condition with 30 μl STP of Kr sample, which lasts for ten hours before the rising pressure due to argon outgassing requires the system to be pumped out and refilled. An analysis run consists of many ~ 40-minute counting cycles: 10 minutes for counting $^{85}$Kr followed by 30 minutes for counting $^{81}$Kr. For a modern atmospheric Kr sample, the typical count rate is 12 hr$^{-1}$ for $^{81}$Kr and 240 hr$^{-1}$ for $^{85}$Kr.

When the rf-discharge is on, a small fraction of the Kr atoms are ionized and imbedded into the surrounding surfaces. Meanwhile, the previously embedded Kr atoms are released into the vacuum. Memory effect arises as the embedded Kr atoms form a source of cross-sample contamination. This effect is mitigated by flushing the system with a discharge of pure $N_2$ or Ar between samples. After 14 hours of flushing, the average amount of residual Kr present during a ten-hour measurement is $(4 \pm 2)\%$ of the Kr sample in the system. It can be reduced further down to $(3 \pm 1.5)\%$ with 36 hours of flushing. This effect should be corrected for each measurement, and will ultimately limit the minimum detection level to perhaps 1% of the modern level.

**Results**

The results of both ATTA and LLC analyses are listed in Table 1. The amount of the samples consumed in ATTA measurements are 50 - 178 µl STP of Kr and the LLC measurements typically consumed 50 µl STP of Kr. Each ATTA measurement took 15 - 20 hours. The error of ATTA measurement of sample 5, a pre-bomb sample free of $^{85}Kr$, is dominated by the error in the correction for the memory effect. The error of the other ATTA measurements is a combination of a dominant statistical error of $^{81}Kr$ counts ($\sim$ 100), a statistical error of $^{85}Kr$ counts, a 5% systematic error due to the uncertainties in laser frequency settings, and an error in the correction for memory effect. The errors of LLC measurements are dominated by the statistical error of $^{85}Kr$ counts. As shown in Fig. 4, the $^{85}Kr/^{81}Kr$ ratios measured with ATTA are indeed proportional to the $^{85}Kr/Kr$ ratios measured with LLC; the best fit, with a reduced-$\chi^2 = 0.6$, indicates that

$[^{85}Kr/^{81}Kr]_{ATTA} = (0.906 \pm 0.040) \times 10^{12} \times [^{85}Kr/Kr]_{LLC}$.

This proportionality is robust as the ten measurements were carried out over a period of six months interspersed with major system changes, including a laser replacement and numerous changes in optical alignments.

Based on these measurements, we derive that for modern atmospheric samples

$[^{81}Kr/Kr]_{modern} = (1.10 \pm 0.05) \times 10^{-12}$

as measured with this method. This value should not yet be taken as the true isotopic abundance of $^{81}Kr$ because ATTA may possess isotope-dependent bias factors that cause it to count one isotope somewhat more efficiently than the other. The weighted mean of three previous measurements on the atmospheric $^{81}Kr/Kr$ measured with LLC (Loosli and Oeschger, 1969; Kuzminov and Pomansky, 1980) and AMS (Collon et al., 1997) is $(0.466 \pm 0.026) \times 10^{-12}$. Even though our measurement is not yet calibrated, we find the disagreement with the previous measurements intriguing, and plan to calibrate our measurement against mass spectrometry in the near future. For $^{81}Kr$ dating, however, the absolute $^{81}Kr/Kr$ ratio is not needed; it is the ratio of $^{81}Kr/Kr$ ratios between the sample and the atmosphere that determines the age.

A detailed analysis of the $^{81}Kr$ budget in the atmosphere concluded that the amount of $^{81}Kr$ released by human activity such as nuclear bomb tests, nuclear fuel reprocessing, and nuclear medicine, is four orders of magnitude less than that produced by cosmic rays in the atmosphere (Collon et al., 1999). A previous measurement by AMS confirmed that the $^{81}Kr/Kr$ ratios of a post-bomb and a pre-bomb atmospheric sample are equal within the ±30% measurement error (Collon et al., 1999). In our work, based on the measurements of post-bomb samples 1, 2, 3, 6 and 9, we conclude that

$[^{81}Kr/Kr]_{post-bomb} = (1.07 \pm 0.06) \times 10^{-12}$;

and based on the measurements of pre-bomb samples 8 and 10, we conclude that

$[^{81}Kr/Kr]_{pre-bomb} = (1.05 \pm 0.08) \times 10^{-12}$.

No changes in the atmospheric $^{81}Kr/Kr$ ratio are observed at the ±8% precision level.

In conclusion, we have demonstrated a method, combining ATTA and LLC, having potential practical applications such as dating old groundwater. Incremental improvements on both the efficiency and counting rate are possible with more laser power and by implementing a more sophisticated transverse cooling scheme. Dramatic improvements may be realized by producing cold metastable Kr atoms with a photon-excitation scheme (Young et al., 2002).

**Acknowledgments.** This work is supported by the U.S. Department of Energy, Nuclear Physics Division, under contract W-31-109-ENG-38, and by the U.S. National Science Foundation grant EAR-0126297. L.Y. is supported by U.S. DOE, Office of Basic Energy Sciences. The work at Bern is supported by the Swiss National Science Foundation and by the University of Bern.

References

Chen, C.-Y., Y.M. Li, K. Bailey, T. O'Connor, L. Young, and Z.-T. Lu, Ultrasensitive isotope trace analysis with a magneto-optical trap, Science, 286, 1139-1141, 1999

Collon, P., T. Antaya, B. Davids, M. Fauerbach, R. Harkewicz, M. Hellstrom, W. Kutschera, D. Morrissey, R. Pardo, M. Paul, B. Sherrill, and M. Steiner, Measurement of $^{81}$Kr in the atmosphere, Nucl. Instr. Meth., B123, 122-127, 1997.

Collon, P., D. Cole, B. Davids, M. Fauerbach, R. Harkewicz, W. Kutschera, D.J. Morrissey, R.C. Pardo, M. Paul, B.M. Sherrill, and M. Steiner, Measurement of the long-lived radionuclide $^{81}$Kr in pre-nuclear and present-day atmospheric krypton, Radiochim. Acta, 85, 13-19, 1999

Collon, P., W. Kutschera, H.H. Loosli, B.E. Lehmann, R. Purtschert, A. Love, L. Sampson, D. Anthony, D. Cole, B. Davids, D.J. Morrissey, B.M. Sherrill, M. Steiner, R.C. Pardo, and M. Paul, $^{81}$Kr in the Great Artesian Basin, Earth Planet. Sci. Lett., 182, 103-113, 2000.

Forster, M., P. Maier, and H. H. Loosli. 1992. Current techniques for measuring the activity of $^{37}$Ar and $^{39}$Ar in the environment. in *Isotopes of Noble Gases as Tracers in Environmental Studies*, pp. 63-72, IAEA, Vienna, 1992.

Kuzminov, V.V., and A.A. Pomansky, New measurement of the $^{81}$Kr atmospheric abundance, Radiocarbon, 22, 311-317, 1980.

Loosli, H.H. and H. Oeschger, $^{37}$Ar and $^{81}$Kr in the atmosphere, Earth Planet. Sci. Lett., 7, 67-71, 1969.

Loosli, H.H., Applications of $^{37}$Ar, $^{39}$Ar and $^{85}$Kr in hydrology, oceanography and atmospheric studies, in *Isotopes of Noble Gases as Tracers in Environmental Studies*, pp. 73-86, IAEA, Vienna, 1992.

Loosli, H. H., B. E. Lehmann, and W. M. J. Smethie. Noble gas radioisotopes: $^{37}$Ar, $^{85}$Kr, $^{39}$Ar, $^{81}$Kr. Pages in P. Cook and A. L. Herczeg, editors. *Environmental Tracers in subsurface Hydrology*. Kluwer Academic Publishers. 379-397, 2000.

Lehmann B.E., H.H. Loosli, D. Rauber, N. Thopnnard, R.D. Willis, $^{81}$Kr and $^{85}$Kr in groundwater, Milk River Aquifer, Canada, Applied Geochemistry, Vol. 6, No.4, 425-434, 1991.

Lehmann, B. E., A. Love, R. Purtschert, P. Collon, H. Loosli, W. Kutschera, U. Beyerle, W. Aeschbach Hertig, R. Kipfer, S. K. Frape, A. L. Herczeg, J. Moran, I. Tolstikhin, and M. Groening. 2003. A comparison of groundwater dating with $^{81}$Kr, $^{36}$Cl and $^{4}$He in 4 wells of the Great Artesian Basin, Australia. Earth and Planetary Science Letters, 212, 237-250, 2003.

Lu, Z.-T. and K.D.A. Wendt, Laser-based methods for ultrasensitive trace-isotope analyses, Rev. Sci. Instr. 74, 1169-1179, 2003.

Young, L., D. Yang, and W. Dunford, Optical production of metastable rare gases, J. Phys. B35, 2985-2988, 2002.

Mailing address: K. Bailey, X. Du, Z.-T. Lu, P. Mueller, T. P. O'Connor, L. Young, *Physics Division, Argonne National Laboratory, Argonne, IL 60439, USA. (lu@anl.gov)*

R. Purtschert, B. Lehmann, R. Lorenzo, *Climate and Environmental Physics, Physics Institute, University of Bern, Sidlerstr. 5, 3012 Bern, Switzerland. (purtschert@climate.unibe.ch)*

N. C. Sturchio, *Department of Earth and Environmental Sciences, University of Illinois at Chicago, Chicago, Illinois 60637, USA. (sturchio@uic.edu)*

8**Fig. 1.** Schematic of a shielded gas proportional counter used in LLC measurements.

**Fig. 2.** Schematic of a second-generation ATTA system.

**Fig. 3.** Signal of a single $^{81}$Kr atom. The signal is due to fluorescence of the trapped atom; the background is due to photons scattered off the surrounding walls of the vacuum chamber.

**Fig. 4.** Proportional correlation between the $^{85}$Kr/$^{81}$Kr ratios measured with ATTA and the $^{85}$Kr/Kr ratios measured with LLC.

**Table 1.** Results of ATTA and LLC analyses.

| No. | Sample Description | Size# (μl STP) | $^{85}$Kr/$^{81}$Kr ATTA | $^{85}$Kr/Kr×10$^{12}$ LLC |
|---|---|---|---|---|
| 1 | Postbomb commercial | 178 | 3.64 ± 0.37 | 4.37 ± 0.49 |
| 2 | Postbomb commercial | n.d. | 16.3 ± 1.9 | 17.9 ± 0.97 |
| 3 | Postbomb lab-prepared | 75 | 20.0 ± 3.0 | 19.7 ± 1.5 |
| 4 | Mixture | 56 | 6.32 ± 0.91 | 8.41 ± 0.33 |
| 5 | Prebomb commercial | n.d. | 0.21 ± 0.26 | 0.00±0.10 |
| 6 | Postbomb commercial | 71 | 19.5 ± 2.1 | 22.5 ± 1.2 |
| 7 | Mixture | 61 | 12.9 ± 1.8 | 13.7 ± 0.91 |
| 8 | Prebomb + spike | 171 | 25.4 ± 2.7 | 24.6 ± 1.5 |
| 9 | Postbomb commercial | 50 | 22.9 ± 2.4 | 22.5 ± 1.2 |
| 10 | Prebomb + spike | 53 | 19.3 ± 2.0 | 21.9 ± 0.61 |

**#** Sample size is the size consumed in the ATTA measurement. The sizes of samples 2 and 5 were not measured.
**\*** The pre-bomb sample was collected from air sometime between 1930 and 1940 in Germany. Among the three pre-bomb samples, sample 5 was not spiked with $^{85}$Kr, samples 8 and 10 were spiked.
**§** Samples 1, 2, 6, and 9 originate from commercial gas processing companies; sample 3 was extracted from 140 liters of modern air; samples 4 and 7 are mixtures between post-bomb and pre-bomb Kr.

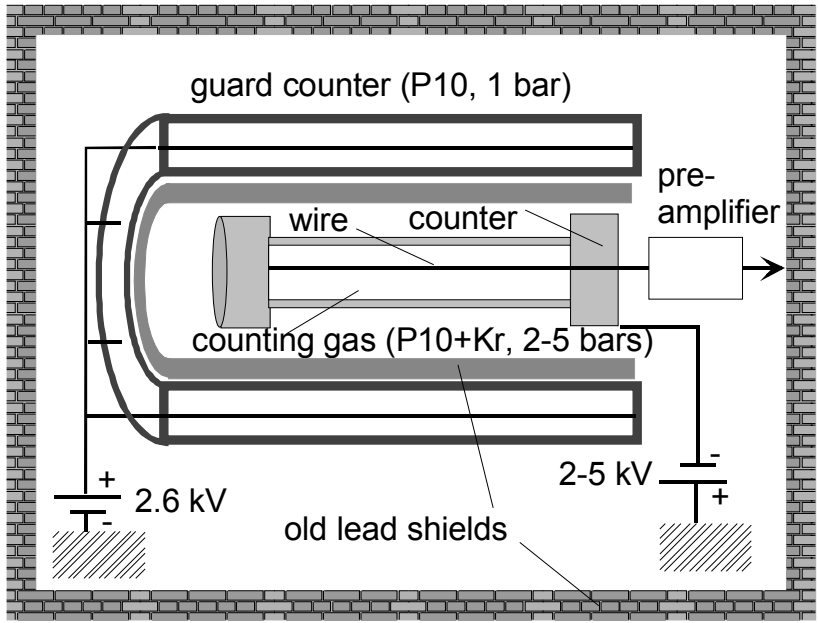

**Fig. 1**

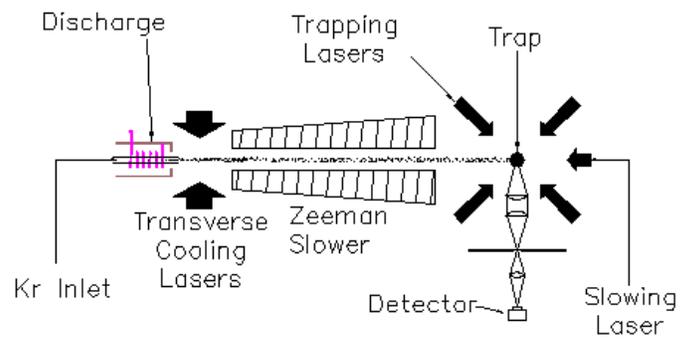

**Fig. 2**

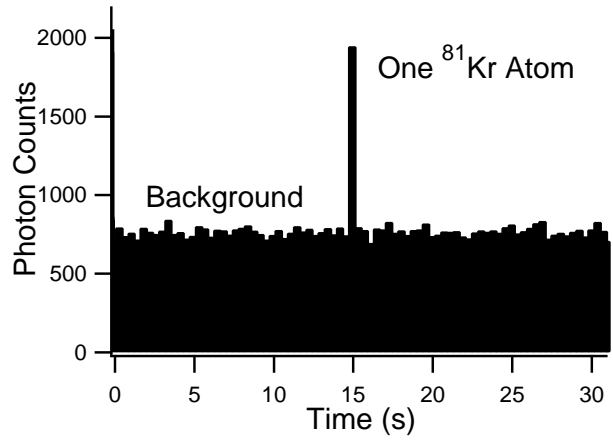

**Fig. 3**

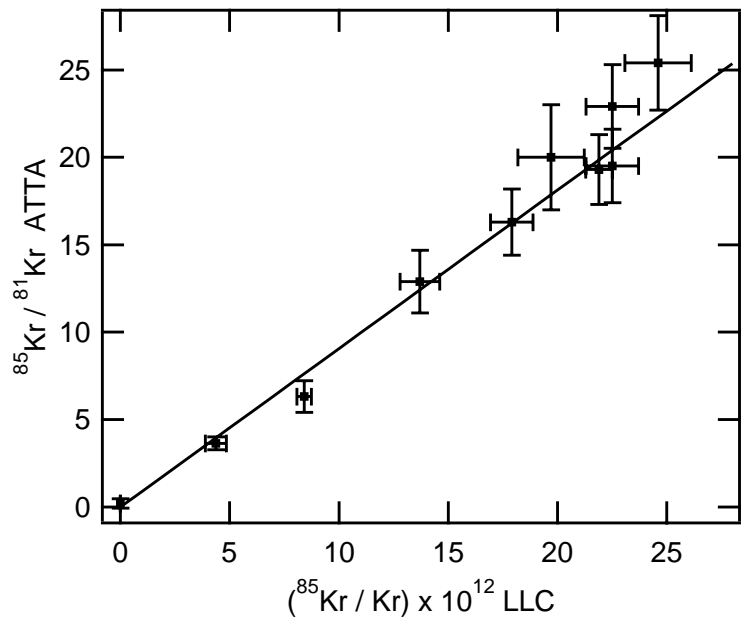

**Fig. 4**